\documentclass[conference,colorlinks=true, linkcolor=blue, citecolor=blue]{IEEEtran}
\IEEEoverridecommandlockouts
\usepackage[colorinlistoftodos,prependcaption,textsize=tiny]{todonotes}
\usepackage{cite}  

\usepackage{url}   
\usepackage[most]{tcolorbox}
\usepackage{amsmath,amssymb,amsfonts}
\usepackage{algorithmic}
\usepackage{graphicx}
\usepackage{textcomp}
\usepackage{ragged2e}
\usepackage{xcolor}
\usepackage{orcidlink}
\usepackage{booktabs} 
\usepackage{array} 
\def\BibTeX{{\rm B\kern-.05em{\sc i\kern-.025em b}\kern-.08em
    T\kern-.1667em\lower.7ex\hbox{E}\kern-.125emX}}
\renewcommand{\footnoterule}
{
  \kern -3pt
  \hrule width \columnwidth
  \kern 2.6pt
}

\begin{document}

\title{Probabilistic Link Budget Analysis for Low Earth Orbit Satellites in the Optical Regime}


\author{
\IEEEauthorblockN{Dhruv Shivkant\orcidlink{0009-0009-3533-8849}\textsuperscript{1}}\thanks{This work is funded by QOSMIC Satellite Systems}
\IEEEauthorblockA{\textsuperscript{1}\textit{Indian Institute of Science}\\
Bangalore, India \\
dhruvsp@iisc.ac.in}
\and
\IEEEauthorblockN{Shreyaans Jain\orcidlink{0009-0004-3669-884X}\textsuperscript{1,2}}
\IEEEauthorblockA{\textsuperscript{1}\textit{Indian Institute of Science} \\
Bangalore, India \\
\textsuperscript{2}\textit{QOSMIC} \\
Bangalore, India \\
shreyaansj@iisc.ac.in\\
shrey@qosmic.space}
\and
\IEEEauthorblockN{Rohit K Ramakrishnan\orcidlink{0000-0001-5936-2346}\textsuperscript{1,2}}
\IEEEauthorblockA{\textsuperscript{1}\textit{Indian Institute of Science} \\
Bangalore, India  \\
\textsuperscript{2}\textit{QOSMIC} \\
Bangalore, India  \\
rohithkr@iisc.ac.in\\
rohit@qosmic.space}

}

\maketitle

\begin{abstract}
Low Earth Orbit (LEO) Optical satellite communication systems face performance challenges due to atmospheric effects such as scintillation, turbulence, wavefront distortion, beam spread, and jitter. This paper presents a comprehensive mathematical model to characterise these effects and their impact on signal propagation. We develop a methodology for dynamically calculating link budgets at any location and time by integrating these models into a probabilistic framework. The approach accounts for spatial and temporal variations in atmospheric conditions, enabling accurate estimation of link loss probabilities. Simulations validate the model’s accuracy and applicability to real-world LEO satellite systems. This work offers a robust tool for optimising link performance and enhancing the reliability of satellite networks, providing valuable insights for system designers and operators.
\end{abstract}


\section{Introduction}
Low Earth Orbit (LEO) satellites are critical infrastructure for Earth observation and global communication networks. Recent advances in data transmission requirements have intensified the demand for faster and more efficient communication systems. Optical communication systems, which offer superior bandwidth capabilities and reduced power consumption compared to traditional radio frequency systems, present compelling advantages for addressing these requirements.
\par Link budget analysis is essential for quantifying various atmospheric channel impairments, including molecular scattering and absorption, scintillation, pointing-induced jitter, beam divergence, and wavefront distortion. These phenomena collectively degrade the performance and reliability of optical laser communication links. Conventional deterministic link budget methodologies fail to adequately account for these atmospheric effects' inherently stochastic characteristics, resulting in suboptimal and potentially unreliable system designs.
\par This paper proposes a probabilistic framework that incorporates the stochastic nature of atmospheric perturbations and optical system parameters, thereby enabling the development of more accurate and robust communication link designs. By accounting for the statistical variability inherent in these systems, the proposed approach facilitates improved performance prediction and enhanced design reliability for LEO satellite optical communication applications for downlink parameters.

\section{Link Budget Equation}
Link Budget for optical communication is a power relation equation. Our work presents the probabilistic analysis (Clements et al. \cite{clements2016nanosatellite}) of this equation :
\begin{equation}
P_{Rx} = P_{Tx}L_{path}\eta_{Rx}\eta_{Tx}G_{Rx}G_{Tx}\eta_{atm}^{static}\eta_{atm}^{dynamic}
\label{eq:link_budget_equation}
\end{equation}
Here,
\begin{itemize}
    \item $P_{Rx}$ is power detected at the receiver 
    \item $P_{Tx}$ is transmitted power
    \item $L_{path}$ is free space propagation loss
    \item $\eta_{Rx}$ is optical efficiency of receiver
    \item $\eta_{Tx}$ is optical efficiency of transmitter
    \item $G_{Rx}$ is gain of receiver 
    \item $G_{Tx} $ is gain of transmitter 
    \item $\eta_{atm}^{static}$ is atmospheric transmittance
    \item $\eta_{atm}^{dynamic}$ accounts for the attenuation due to atmospheric turbulence.
    
\end{itemize}
For our analysis, we convert the above linear equation to logarithmic terms,

\begin{equation}
P_{Rx}^{dB} = P_{Tx}^{dB}+L_{path}^{dB}+L_{Rx}^{dB}+L_{Tx}^{dB}+G_{Rx}^{dB}+G_{Tx}^{dB}+{L_{atm}^{static}}^{dB}+{L_{atm}^{dynamic}}^{dB}
\end{equation}
Here, the superscript $(.)^{dB}$ refers to the logarithmic analogue of the linear term and is calculated as $10\log_{10}(.)  \quad (in \ dB)$.
$L_{Rx}^{dB}, L_{Tx}^{dB}
$ corresponds to losses in the receiver and transmitter telescopes, respectively.
\par We categorize the link parameters into static and dynamic (i.e. time-invariant and time-variant), and carry out further analysis as follows.

\section{Static components of link budget}
\subsection{Free space propagation loss}
The free space propagation loss is the most significant loss encountered while establishing an optical link. It depends on the wavelength and the distance between the transmitter and receiver.
For an isotropic radiator, its magnitude is given by Friis' transmission loss equation by Johnson et al. \cite{johnson1984antenna} as
\begin{equation}
L_{path}^{dB} = 20\times\log_{10}(\lambda/4\pi L)
\label{eq:path_loss}
\end{equation}
where $\lambda$ is the wavelength of the transmitted laser and L is the distance between the transmitter and receiver. In the general case when satellite is not at the zenith, we can use the following formula to find the value of L;
\[L = \sqrt{R_{GS}^2 [\sin(\varepsilon)]^2 + 2(\Delta H)R_{GS} + \Delta H^2} - R_{GS} \sin(\varepsilon)\]
where,
\begin{itemize}
    \item $\Delta H =H_0 - H_{GS}$ is difference between altitude of satellite and ground station.
    \item $R_{GS} = R_E + H_{GS}$ is distance of ground station from the centre of the earth
    \item $\varepsilon$ is angle of elevation.
\begin{figure}
    \centering
    \includegraphics[width=1\linewidth]{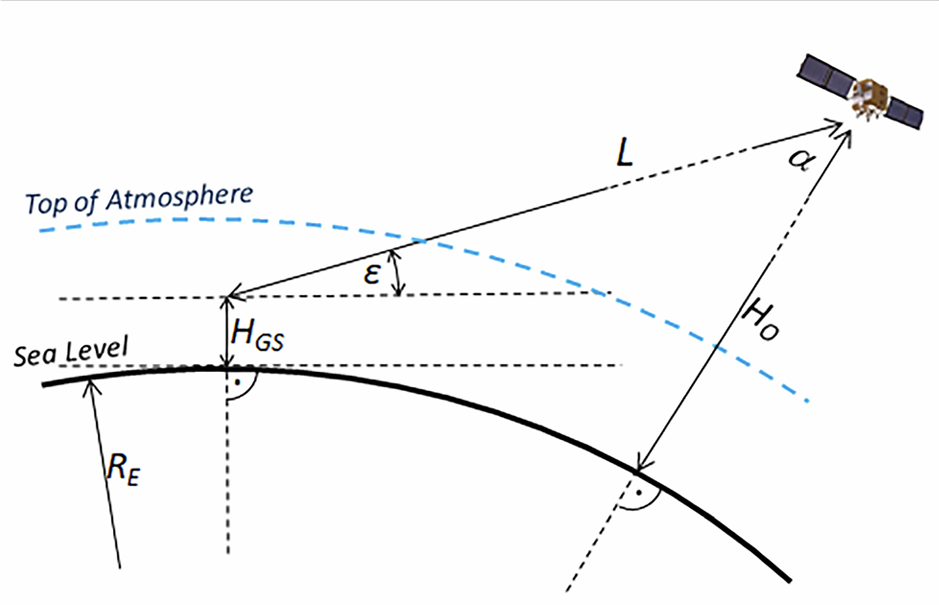}
    \caption{Angles and distances in general Earth-GS-sat triangle (Credit: Giggenbach et al. \cite{giggenbach2023link}}
    \label{fig:placeholder}
\end{figure}
\end{itemize}
\subsection{Transmitter and Receiver optics}
The optical efficiency for a Cassegrain-type transmitter telescope architecture taking into account central obstruction due to secondary mirror is given by Klein et al. \cite{klein1974optical}:
\begin{equation}
\eta_{Tx} = \frac{2}{\alpha^2} \left( e^{-\alpha^2} - e^{-\alpha^2 \gamma^2} \right)^2
\end{equation}
where $\alpha = R/\omega$. $R$ is the primary radius of the telescope, and $\omega$ is the beam radius at the transmitter end. $\gamma = b/R$, corresponds to the ratio of the telescope's secondary radius to primary radius. Thus, the transmitter loss is found as
\begin{equation}
L_{Tx}^{dB} = 10\log_{10}(\eta_{Tx})
\end{equation}
In addition to this, there may be additional implementation losses ranging from 1 - 4 dB depending on coupling efficiency, lens assembly, etc.
The optical efficiency of a Cassegrain-type receiver telescope is given by Degnan et al. \cite{Degnan:74}:
\begin{equation}
\eta_{Rx} =  (1 - \gamma 
^2)\times \eta_{detector}
\end{equation}
Where $\gamma = b/R$, corresponds to the ratio of the receiver telescope's secondary radius to primary radius and $\eta_{detector}$ is the efficiency of the detector mechanism. The receiver loss is then calculated as
\begin{equation}
    L_{Rx}^{dB} = 10\log_{10}(\eta_{Rx})
\end{equation}
The gain of the receiver is given by Behera et al. \cite{behera2024estimating}
\begin{equation}
G_{Rx} = 4\pi A_r/\lambda^2
\label{eq:G_rx}
\end{equation}
where $A_r$ (in $m^2$) is the area of the receiver.\\
The gain of the transmitter is given by Giggenbach et al. \cite{giggenbach2023link}
\begin{equation}
G_{Tx} = 8/(\theta_B)^2
\label{eq:G_tx}
\end{equation}
where $\theta_B$ is the beam's half-divergence angle (in rad).
\subsection{Atmospheric absorption and scattering}
The atmosphere results in the power loss due to absorption and scattering by various constituents of the atmosphere, such as water vapour, gases, dust particles, etc. This loss is primarily static and is dictated by Beer-Lambert's law \cite{thomas2006optical}, which models this loss as;
\begin{equation}
I(L) = I_0 \cdot e^{-\beta L}
\end{equation}

where:
\begin{itemize}
    \item $I(L)$ (in $W/m^2$) is the transmitted intensity after propagating distance $L$
    \item $I_0$ (in $W/m^2$) is the initial incident intensity
    \item $\beta$ is the extinction coefficient (m$^{-1}$) which accounts for the absorption and scattering effects
    \item $L$ is the propagation path length (km) containing atmosphere.
\end{itemize}

The atmospheric transmittance, $T$ can be expressed as:
\begin{equation}
T_{atm} = \frac{I(L)}{I_0} = e^{-\beta L}
\end{equation}
Alternatively, if atmospheric transmittance at zenith ($T_{zen}$) is known, then for angle of elevation $\varepsilon$, the atmospheric transmittance is given by,
\[
T_{atm,\varepsilon} = T_{zen}^{1/sin(\varepsilon)}
\]
The atmospheric transmittance was calculated using MODTRAN, a popular simulation software for modeling the atmosphere from 100,000 nm (Far-IR) to 200 nm (UV). Using this, we can get the static atmospheric loss for a particular wavelength as follows.

\begin{equation}
{L_{atm}^{static}}^{dB} = 10\log_{10}(T_{atm}) = -4.342\beta L
\end{equation}

\section{Dynamic components of link budget}
\subsection*{Modeling of atmospheric turbulence}
 The atmospheric turbulence is a dynamic parameter that affects optical communication. Using the refractive index structure parameter and atmospheric coherence length,  we can further analyse turbulence-induced effects - scintillation, jitter, beam spread and wavefront distortion.
\par The refractive index structure parameter ($C_n^2(h)$) \cite{andrews2019field} is the most significant parameter in modeling turbulence. Direct calculation of $C_n^2(h)$ involves considering temperature gradient, wind speeds and several other meteorological factors, which is often computationally expensive. Hence, we use parametric models to estimate it. One of the more popularly used parametric models is the Huffnagel-Valley model\cite{andrews2019field}  given as 
\begin{equation}
\begin{aligned}
C_n^2(h) = &\: 0.00594\left(\frac{v}{27}\right)^2(10^{-5}h)e^{-10^{-3}h} \\
            &+ 2.7 \times 10^{-16}e^{-\frac{h}{1500}} \\
            &+ A_0 e^{-10^{-2}h}
\end{aligned}
\end{equation}
Here, $h$ (in $m$) is the altitude above sea level, $A_0$ ($m^{-2/3}$) defines the turbulence strength at ground level and $v$ ($ms^{-1}$) is the rms wind speed at high altitude. These values change with time and are hence dynamic. For our calculations, we have considered standard values for these as $1.7\times 10^{-14}\ m^{-2/3}$ and $21\ ms^{-1}$ respectively \cite{andrews2019field}.
\\
\par The atmospheric coherence length ($r_0$) is a crucial parameter in describing the turbulence. It can be interpreted as the diameter of an equivalent aperture where the root-mean-square phase variation 
is approximately 1 rad. Using the HV parametric model, we can calculate the atmospheric coherence length (fried parameter) as

\begin{equation}
r_0 = \{0.42k^2 \sec \theta \int_{h_0}^Z C_n^2(h) \, dh\}^{-3/5}
\label{eq:r_0}
\end{equation}
where:
\begin{itemize}
    \item $r_0$ is the atmospheric coherence length (m)
    \item $k$ is the wave number (m$^{-1}$)
    \item $\theta$ is the zenith angle (rad)
    \item $h_0$ is the initial altitude (m)
    \item $Z = h_0 + L$ is the final altitude (m) (L is the distance between the satellite and the ground station)
    \item $C_n^2(h)$ is the refractive index structure parameter (m$^{-2/3}$)
\end{itemize}

We now describe four significant types of turbulence-induced link attenuators in optical communications: Scintillation, Jitter, Beam Spread and Wavefront Distortion.

\subsection{Scintillation}
Scintillation refers to the fluctuation of signal irradiance due to air turbulence. A stochastic modelling of this involves a random variable that follows a log-normal distribution with unit mean and variance \cite{piazzolla_atmospheric}
\begin{equation}
\sigma^2 = exp(\sigma_I^2)-1
\end{equation}
Here $\sigma_I^2$ is called the scintillation index. The formulas provided in Table \ref{tab:scintillation index formulas} can be found for a particular case.
\begin{table}
\caption{Scintillation index formulas  \cite{piazzolla_atmospheric}}
\label{tab:scintillation index formulas}
\centering
\footnotesize
\setlength{\tabcolsep}{2pt}
\begin{tabular}{@{}l>{ $ \displaystyle }l<{ $ }@{}}
\toprule
\textbf{Wave Type} & \textbf{Formula for $\sigma_I^2$ ($\chi = 2.24k^{7/6}\sec^{11/6}\theta$)} \\ 
\midrule
Downlink Plane & 
\chi \int_{h_0}^{h_0+L} C_n^2(h)(h-h_0)^{5/6} dh \\

Downlink Spherical & 
\chi \int_{h_0}^{h_0+L} C_n^2(h) (h-h_0)^{5/6} \\
 & \quad \times \left(\!1-\frac{h-h_0}{L}\!\right)^{\!\!5/6} dh \\

Uplink Plane & 
\chi \int_{h_0}^{h_0+L} C_n^2(h)(L+h_0-h)^{5/6} dh \\

Uplink Spherical & 
\chi \int_{h_0}^{h_0+L} C_n^2(h)(L+h_0-h)^{5/6} \\
 & \quad \times \left(\!1-\frac{L+h_0-h}{L}\!\right)^{\!\!5/6} dh \\
\bottomrule
\end{tabular}
\end{table}

 For the downlink of a planar wavefront, it is given as
\begin{equation}
\sigma_I^2 = 2.24k^{7/6}\sec^{11/6}\theta \int_{h_0}^{h_0+L} C_n^2(h)(h-h_0)^{5/6} \, dh 
\label{eq:scintillation_index_downlink_planar}
\end{equation}
\subsubsection*{Aperture Averaging}
The equations in Table \ref{tab:scintillation index formulas} and eq.\ref{eq:scintillation_index_downlink_planar} are based on the assumption that the receiver is a point object. However, in reality, this is not the case. Hence, the scintillation effect is offset by averaging across the surface of the receiver. This effect can be quantified by the aperture averaging factor $A_v$ \cite{piazzolla_atmospheric}, which is defined as the ratio between $\sigma_I^2(D)$ (normalised flux variance of the collection aperture of diameter D) and the scintillation index $\sigma_I^2$.
Thus,
\begin{equation}
A_v = \sigma_I^2(D)/ \sigma_I^2 
< 1
\end{equation}
An estimate of $A_v$ for the case of downlink is provided by Piazzolla \cite{piazzolla_atmospheric} as

\begin{equation}
A_v = \left[1 + 1.1 \left(\frac{D^2}{\lambda h_s \cos(\theta)}\right)^{7/6}\right]^{-1} 
\end{equation}
Here, $h_s$ is the scale height given for the downlink of plane waves by
\begin{equation}
h_s = \left[\frac{\int_{h_0}^{h_0+L} C_n^2(h)(h - h_0)^2 \, dh}{\int_{h_0}^{h_0+L} C_n^2(h)(h - h_0)^{5/6} \, dh}\right]^{6/7}
\label{eq:scale_height}\end{equation}
Taking this into account, the variance of S (a random variable representing scintillation) becomes
\begin{equation}
\sigma^2 = exp(\sigma_I^2 \times A_v) - 1
\label{eq:scintillation_with_aperture_average}\end{equation}

\subsection{Jitter}

Transmitter-induced mechanical jitter can also be modelled as a random variable  \cite{behera2024estimating} following a beta distribution with parameters $(\beta,1)$. Here,
\begin{equation}
\beta = (\theta_B)^2/4\sigma_j^2
\label{eq:beta}\end{equation}
The variance $\sigma_j^2$ is made up of the variance of atmospheric-induced and transmitter-induced pointing errors and can be found by \cite{behera2024estimating}:
\begin{equation}
\sigma_j^2 = 0.182 (D_{Tx}/r_0)^{5/3}(\lambda/D_{Tx})^{2}
\label{eq:sigma_j_square}\end{equation}
$D_{Tx}$ is the diameter of the transmitter and $r_0$ is the fried parameter as described in eq.\ref{eq:r_0}.

\subsection{Beam Spread}
Beam spread refers to broadening the beam size due to propagation through an optical medium, such as the atmosphere. In the absence of air turbulence, the size of a Gaussian beam at the receiver end can be obtained by \cite{hecht2017optics}:
\begin{equation}
w(L) = w_0[1 + (L/Z_r)^2]^{0.5}
\end{equation}
where L is the path range, $w_0$ is the beam waist and $Z_r = \pi w_0^2/\lambda$ is the Rayleigh distance.
\\
\par However, in the presence of turbulence, the variable refractive index of the air causes the beam size to increase further, causing a degradation in the received link quality. This effective beam radius is found to be
\begin{equation}
w_{eff}(L) = w(L)(1+T)^{1/2}
\label{eq:w(L)}\end{equation}
Here, T is representative of the turbulence, and for downlink of planar wavefronts, it is given by Piazzolla \cite{piazzolla_atmospheric} as
\begin{equation}
   T = 4.35\Lambda^{5/6} k^{7/6} L^{-5/6} \sec^{11/6}(\theta) \int_{h_0}^{Z} C_n^2(h) 
   ({h-h_0})^{5/3} dh 
\label{eq:T}\end{equation}

where,
\begin{itemize}
    \item $\Lambda = 2L/k(w(L))^2$  
    \item $w(L)$ is the beam waist at the receiver in the absence of turbulence 
    \item L is the path range
    \item $h_0$ is altitude of ground station and $Z = h_0 + L$ is altitude of satellite
\end{itemize}
The beam spread loss can then be calculated by
\begin{equation}
L_{beam \ spread} = 20\log_{10}(w(L)/w_{eff}(L)) = -10\log_{10}(1+T)
\label{eq:L_beam_spread}\end{equation}
\subsection{Wavefront Distortion}
Atmospheric turbulence causes the atmosphere to behave like a collection of many tiny lenses, which distorts the wavefront of the transmitted laser beam. This wavefront distortion loss is captured by the Strehl Ratio (S) which is given by Andrews et al. \cite{andrews2006strehl}
\begin{equation}
    S = \left(1 + \left(\frac{D_{Tx}}{r_0}\right)^{5/3}\right)^{-6/5}
\label{eq:S}\end{equation}
where,
\begin{itemize}
    \item $0\leq D_{Tx}/r_0 < \infty$
    \item $D_{Tx}$ is diameter of the transmitter telescope
    \item $r_0$ is fried parameter
\end{itemize}
Hence, the loss due to wavefront distortion is
\begin{equation}
    L_{strehl}^{dB} = 10\log_{10}(S)
\label{eq:strehl}\end{equation}
Note that if $1 \gg D_{Tx}/r_0$ then the scintillation effect is negligible, i.e. $S \approx 1$.
\section{Uncertain parameters modeling}
As opposed to traditional deterministic analysis of link budgets, we adopt a probabilistic analysis by considering the uncertainties introduced into the system via transmitter optics, receiver optics and stochastic modeling of scintillation and jitter losses. These are summarized in Table \ref{tab:distributions}.
\begin{table}
\caption{Terms and Distributions}
\label{tab:distributions}
\centering
\footnotesize
\setlength{\tabcolsep}{4pt} 
\begin{tabular}{@{}>{\RaggedRight}p{0.2\columnwidth}>{\RaggedRight}p{0.25\columnwidth}>{\RaggedRight}p{0.45\columnwidth}@{}}
\toprule
\textbf{Term} & \textbf{Distribution} & \textbf{Rationale} \\
\midrule
$L_{Tx}$ & Normal(0,0.5) &  0.5 dB for miscellaneous losses \\
$L_{Rx}$ & Uniform(0.35,0.5) & -1 dB for beamsplitter and -2 dB for miscellaneous losses \\

$L_{scintillation}$ & lognormal(1,$\sigma$) where $\sigma^2 =e^{\sigma_I^2\times A_v} - 1$ & Kiasaleh \cite{kiasaleh1994probability} \\

$L_{jitter}$ & beta($\beta$,1) & Toyoshima et al. \cite{toyoshima2002optimum} \\
\bottomrule
\end{tabular}
\end{table}
\section{Case study 1}
\begin{figure}
    \centering
    \includegraphics[width=1\linewidth]{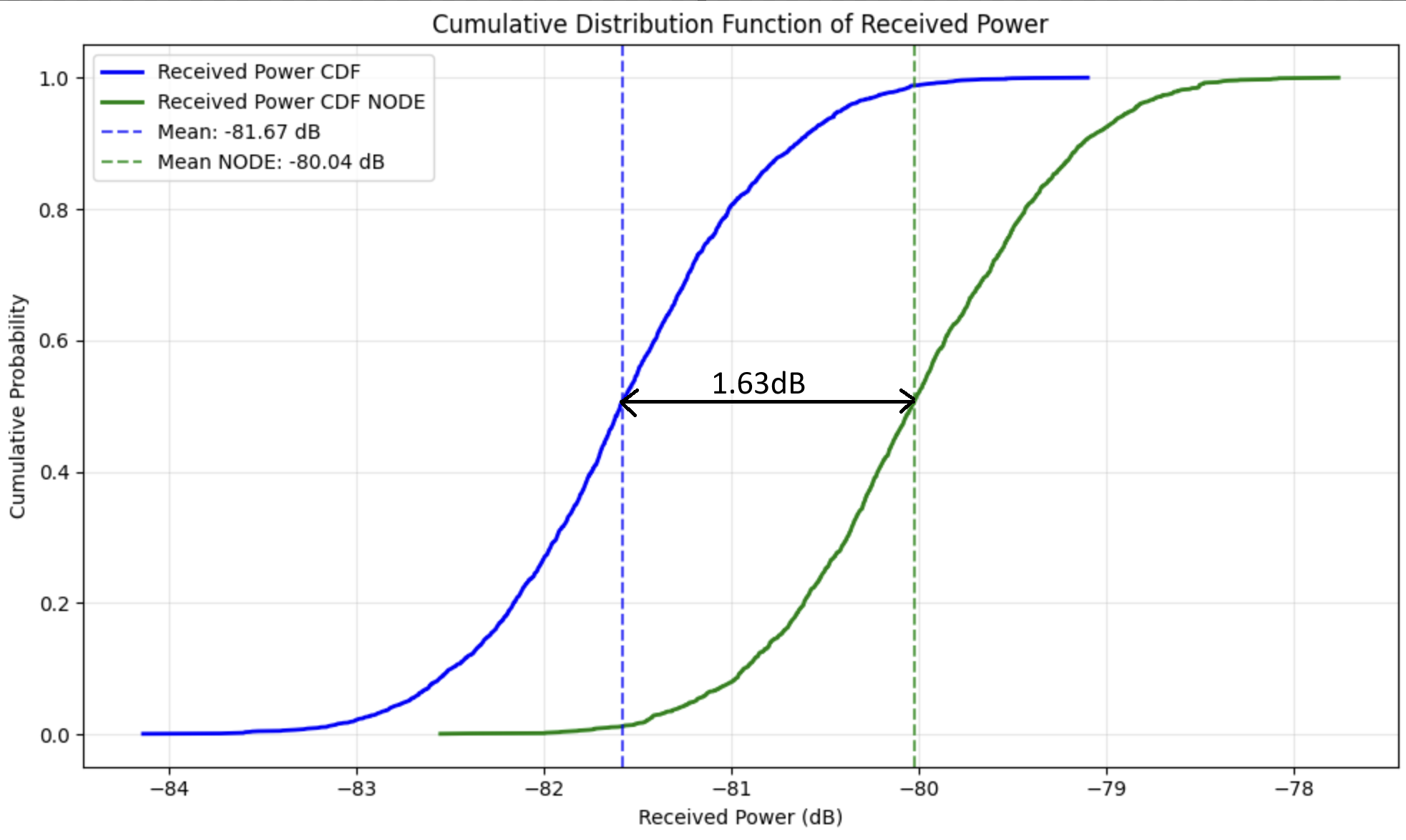}
    \caption{CDF of received power}
    \label{Fig. 1.}
    
\end{figure}

In this section, we validate our model based on the Nanosatellite Optical Downlink Experiment (NODE) (Clements et al. \cite{clements2016nanosatellite}). The details of the downlink setup are as in Table \ref{NODEsetup}. The results of Monte Carlo Analysis (2000 runs) are summarised in Tables \ref{tab:random_stats2} and \ref{tab:link_budget2}, and Fig.~\ref{Fig. 1.}.

\begin{table}
\centering
\caption{Setup in Clements et al. \cite{clements2016nanosatellite} downlink}
\label{NODEsetup}
\begin{tabular}{lrl}
\toprule
\textbf{Item} && \textbf{Value} \\
\midrule
Wavelength (nm) && 1550 \\
Tx power (W) && 0.2 \\
Tx beam waist (mm) ($1/e^2$) && 95 (from CubeSAT specs) \\
Tx optical loss (dB) && -1.5 \\
Propagation distance (km) && 1000\\
Rx diameter (m) && 1 (OCTL)\\
Rx optical loss (dB) && -3\\
\bottomrule
\end{tabular}
\end{table}

\begin{table}
\centering
\caption{Dynamic Terms Statistics}
\label{tab:random_stats2}
\begin{tabular}{lrrrrl}
\toprule
\textbf{Parameter} &&&&& \textbf{Mean Value ({dB})} \\
\midrule
Beam spread loss &&&&& $-6.041\times 10^{-5}$\\
Strehl Loss &&&&&$ -5\times 10^{-5}$\\
Scintillation Loss &&&&& $-0.02$ \\
Jitter Loss &&&&& $-1.69\times 10^{-6}$  \\
\midrule
Atmospheric dynamic loss  &&&&& - 0.02 \\

\bottomrule
\end{tabular}
\end{table}

\begin{table}
\centering
\caption{Optical Link Budget Breakdown}
\label{tab:link_budget2}
\begin{tabular}{lrrrrrl}
\toprule
\textbf{Parameter} &&&&&& \textbf{Value ({dB})} \\
\midrule

Transmitted Power &&&&& &-7 \\
Free Space Path Loss &&&&&& -258.2 \\
Receiver Loss &&&&&& -3.425 \\
Transmitter Loss &&&&&& -1.5 \\
Transmitter Gain &&&&&& 63.39 \\
Receiver Gain &&&&&& 126.1 \\
Atmospheric Static Loss &&&&&& -1 \\
Atmospheric dynamic loss&&&&&& -0.02 \\

\midrule
\textbf{Total Received Power} &&&&&& -81.67\\
\bottomrule
\end{tabular}
\end{table}

\subsubsection*{Findings}
We observe a difference of around 1.63 dB between the mean values of our simulated results and the data provided in Clements et al. \cite{clements2016nanosatellite}. This can be justified by
\begin{itemize}
    \item Difference in transmission gain - The calculated transmission gain using eq.\ref{eq:G_tx} is 63.39 dB, whereas Clements et al. \cite{clements2016nanosatellite} report a value of 65 dB. This discrepancy may be justified by inclusion of the beam quality factor ($M^2$)\cite{Paschotta_2005_m2_factor}. This factor can be attributed to collimated TEM\textsuperscript{00} for laser diodes, with its value ranging from 1.1 - 1.7\cite{ScitecInstrumentsM2}.

    \item We have accounted for an additional atmospheric dynamic loss of 0.02 dB which is primarily due to scintillation.
\end{itemize}
\section{Case Study 2}
For this case study, we consider the configuration as in Behera et al. \cite{behera2024estimating}. We are dealing with a quantum optical downlink of wavelength 1550 nm. Some other supplementary details of the setup are given in Table \ref{HANLEsetup}.

\begin{table}
\centering
\caption{Setup in Behera et al. \cite{behera2024estimating} downlink}
\label{HANLEsetup}
\begin{tabular}{lrrrrrl}
\toprule
\textbf{Item} &&&&&& \textbf{Value} \\
\midrule
Wavelength (nm) &&&&&& 1550 \\
Tx power (W) &&&&&& 1 \\
Tx beam waist (mm) ($1/e^2$) &&&&&& 300 \\
Tx optical loss (dB) &&&&&& -2.2 \\
Propagation distance (km) &&&&&& 500\\
Rx diameter (m) &&&&&& 0.15\\
Rx optical loss (dB) &&&&&& -2.2\\
\bottomrule
\end{tabular}
\end{table}

\begin{figure}
    \centering
    \includegraphics[width=1\linewidth]{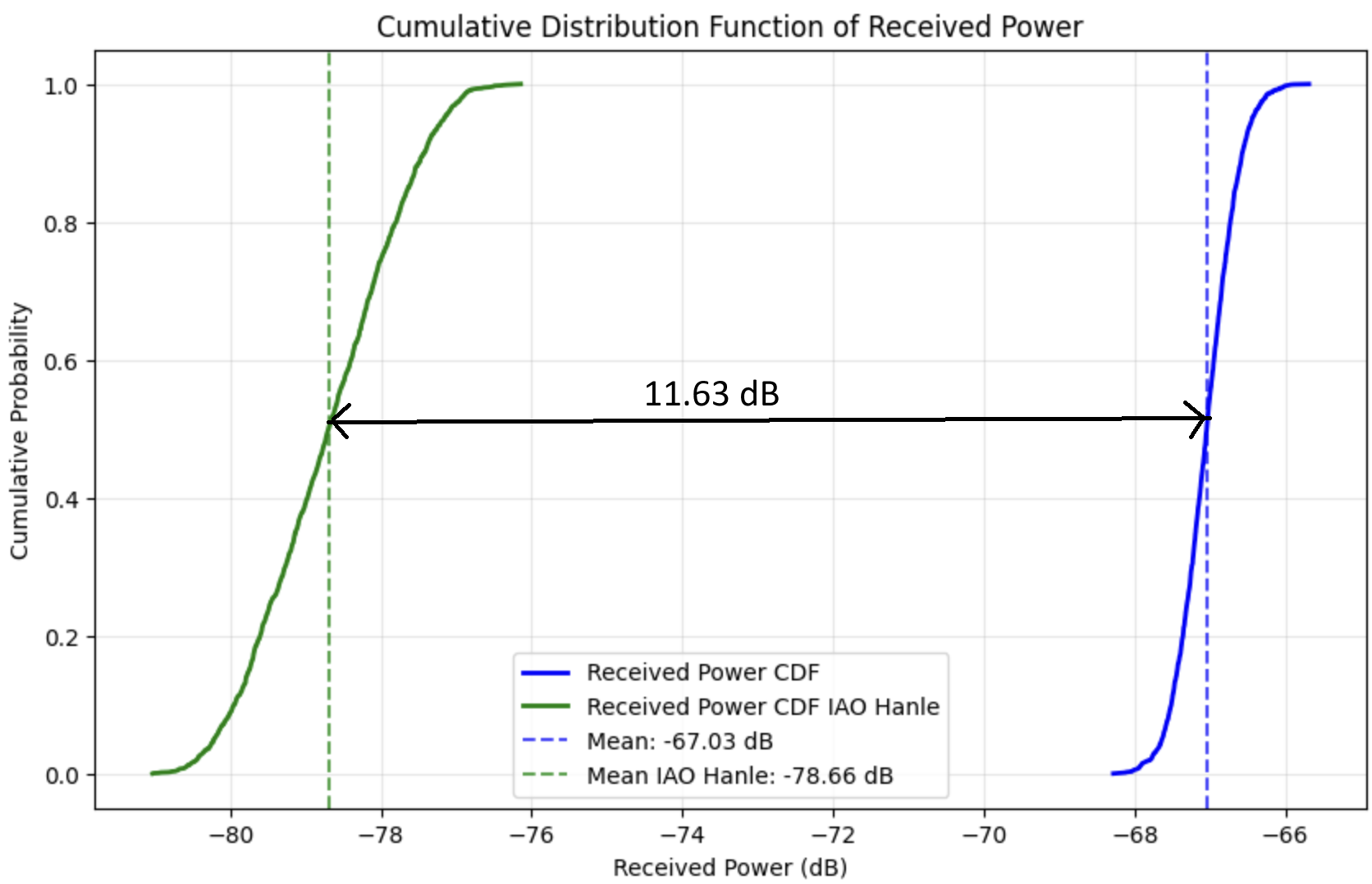}
    \caption{CDF of received power}
    \label{Fig .2.}
    
\end{figure}

We perform a 2000-run Monte Carlo Simulation to estimate the power received at the ground station of IAO Hanle (altitude = 4500 m). The results are as presented in Tables \ref{tab:random_stats} and \ref{tab:link_budget}. Fig.~\ref{Fig .2.} shows the cumulative probability distribution of received power obtained from the simulation.

\begin{table}
\centering
\caption{Dynamic Terms Statistics}
\label{tab:random_stats}
\begin{tabular}{lrrrrl}
\toprule
\textbf{Parameter} &&&&&\textbf{Mean Value ({dB})} \\
\midrule
Beam spread loss &&&& &-0.00081\\
Strehl Loss &&&&& -0.223\\
Scintillation Loss &&&&& -0.0177 \\
Jitter Loss &&&&& -0.000231 \\
\midrule
Atmospheric dynamic loss  &&&&& -0.242 \\

\bottomrule
\end{tabular}
\end{table}
\begin{table}
\centering
\caption{Optical Link Budget Breakdown}
\label{tab:link_budget}
\begin{tabular}{lrrrrrl}
\toprule
\textbf{Parameter} &&&&&&\textbf{Value ({dB})} \\
\midrule

Transmitted Power &&&&&&0.00 \\
Free Space Path Loss &&&&& &-252.153 \\
Receiver Loss &&&&&& -2.20 \\
Transmitter Loss &&&&&& -2.20 \\
Transmitter Gain &&&&&& 81.0721 \\
Receiver Gain &&&&&& 109.654 \\
Atmospheric Static Loss &&&& &&-0.970 \\
Atmospheric dynamic loss&&&& &&-0.242 \\

\midrule
\textbf{Total Received Power} &&&&&& -67.1389\\
\bottomrule
\end{tabular}
\end{table}

\subsubsection*{Findings}
We find that there is a difference of around 11.63 dB in the mean value of received power according to our analysis and the results in Behera et al. \cite{behera2024estimating}. This can be explained by
\begin{itemize}
    \item Difference in path loss - We have obtained path loss as 252.15 dB for 500km downlink as per eq.\ref{eq:path_loss}, whereas Behera et al. \cite{behera2024estimating} quote a loss of 258 dB, which corresponds to a 1000km downlink.
    \item The total optical gain (receiver and transmitter) in our analysis yields 190.72 dB as per eq.\ref{eq:G_tx} and eq.\ref{eq:G_rx}. But Behera et al. \cite{behera2024estimating} use a value of 187.72 dB. The discrepancy may arise from differing analytical approaches. Our study focuses on optical communication, emphasising signal power, unlike the work focusing on quantum communication, where photon intensity is prioritised.
    \item Behera et al. \cite{behera2024estimating} consider a conservative upper bound estimate of the atmospheric turbulence loss as 3 dB. However, our results indicate that for downlink, this value is at 0.242 dB, i.e. much less than the conservative estimate of 3 dB. 
\end{itemize}

\section{Case Study 3}
\begin{figure}[h]
    \centering
    \includegraphics[width=1\linewidth]{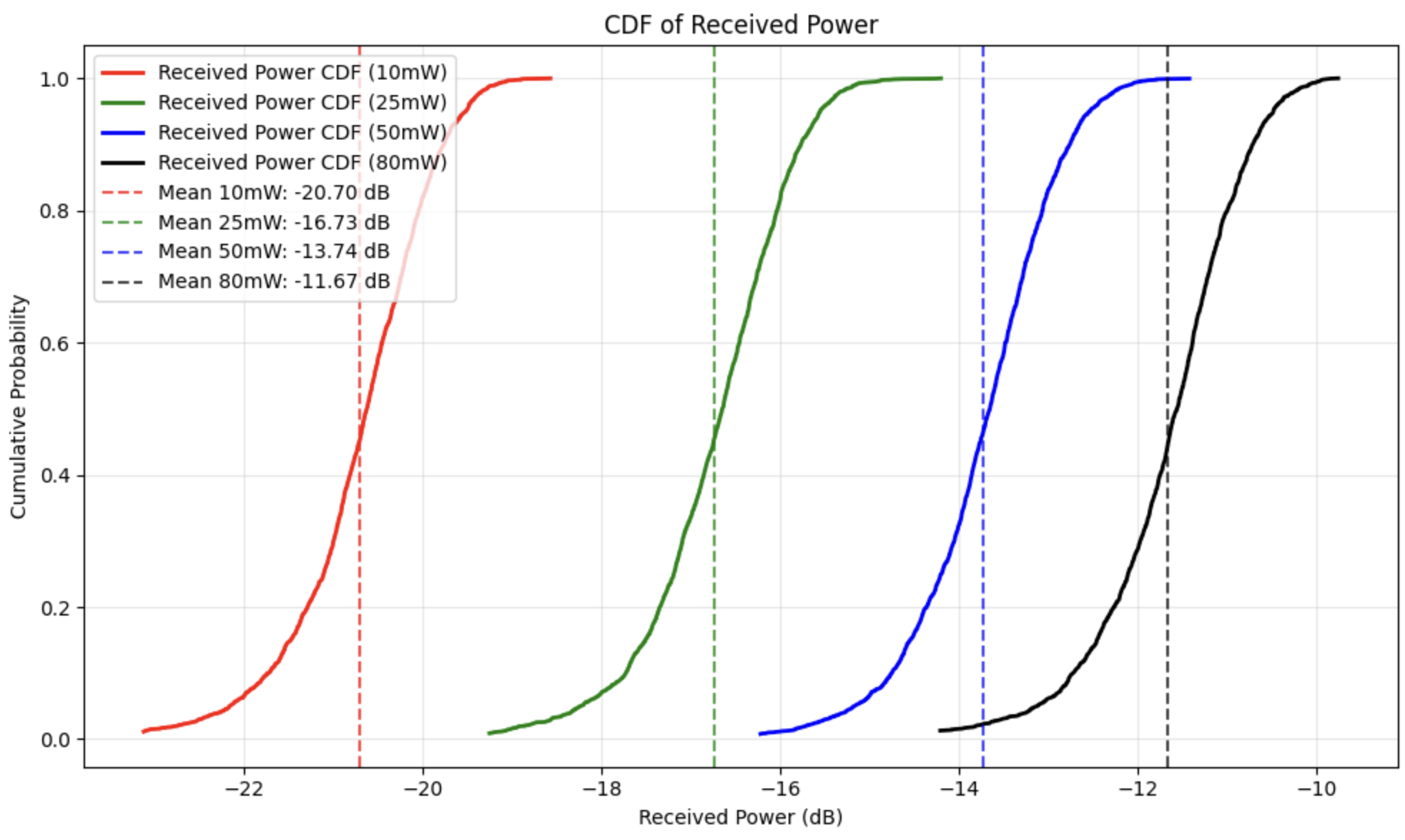}
    \caption{CDF of received power}
    \label{balooncdf}
\end{figure}

We now simulate the results for optical downlink from a helium balloon at an altitude of 35 km above sea level. According to Behera et al. \cite{behera2024estimating} the optimum transmitter radius for downlink with a laser of wavelength 1550 nm is around 14-15 cm. Table \ref{tab:balloon} and Fig.\ref{balooncdf} present the results of link budget simulation carried out for a 15cm transmitter aperture with four different laser powers - 10mW, 25mW, 50mW and 80mW.

\begin{table}
\centering
\caption{$P_{Rx}^{dB}$ for different transmitter powers ($\lambda = 1550 nm)$}
\label{tab:balloon}
\begin{tabular}{lrl}
\toprule
\textbf{Tx Power (dB)} & \textbf{Rx power (dB)} & \textbf{Range of $P_{Rx}^{dB}$ in CDF}\\
\midrule

-20 & -20.70 & -23.21 to -18.20 \\
-16.02 & -16.72 & -19.14 to -14.26 \\
-13.01 & -13.75 & -16.18 to -11.25 \\
-10.96 & -11.66 & -14.08 to -9.22 \\

\bottomrule
\end{tabular}
\end{table}

\section{Conclusion and Future Work}
In this work, we have developed and validated a comprehensive probabilistic framework to analyse the link budget of LEO optical downlinks. Our approach provides an advanced and physically grounded estimation of link performance by combining detailed models for dynamic atmospheric effects like scintillation, jitter, and wavefront distortion with static system parameters. The case studies demonstrate that this probabilistic approach produces more precise, location-specific results than traditional deterministic analyses, especially when measuring atmospheric losses. This methodology offers a powerful tool for developing more reliable and efficient LEO communication systems by enabling a more accurate characterisation of link margins and minimising the over-engineering typically associated with conservative loss estimates.

\textit{Future Work:}
This analysis will be expanded upon in future research by addressing its main shortcomings in two crucial areas. First, uplink transmissions will be added to the scope of the investigation. Until now, we have only examined downlink scenarios, where atmospheric turbulence has a less significant effect. Since the laser must travel through the entire atmosphere before a noticeable increase in beam size, the probabilistic link budget model must be extended to uplink configurations. This increases the link's susceptibility to turbulence-induced losses and pointing errors.

Second, the current scintillation model will be generalised to improve the framework's accuracy and robustness under various atmospheric conditions. The weak turbulence assumption, which is the foundation of the current methodology, is not always valid in practical operational settings. The model will be expanded to include a strong turbulence regime to address this limitation. A more flexible and trustworthy tool for forecasting optical link performance can be produced by incorporating a more thorough analytical framework \cite{piazzolla_atmospheric}.

\bibliographystyle{ieeetr}  
\bibliography{references} 
\end{document}